\documentclass[runningheads]{llncs}
\usepackage{graphicx}
\usepackage{amsmath,amssymb,amsfonts}
\usepackage{algorithmic}
\usepackage{graphicx}
\usepackage{textcomp}
\usepackage{hyperref}
\usepackage{xcolor}
\usepackage{booktabs}
\newcommand{\ra}[1]{\renewcommand{\arraystretch}{#1}}

\begin{document}
\title{A Self-attention Guided Multi-scale Gradient GAN for Diversified X-ray Image Synthesis\thanks{This work is supported by the Munster Technological University's Risam Scholarship Award}}
\author{Muhammad Muneeb Saad\inst{1}\orcidID{0000-0002-0204-0597} \and
Mubashir Husain Rehmani\inst{1}\orcidID{0000-0002-3565-7390} \and
Ruairi O'Reilly\inst{1}\orcidID{0000-0001-7990-3461}} 

\authorrunning{Muhammad Muneeb Saad. et al.}
\institute{Munster Technological University Cork, Ireland \\
\email{muhammad.saad@mycit.ie, mubashir.rehmani@mtu.ie, and ruairi.oreilly@mtu.ie}
}
\maketitle              % typeset the header of the contribution
\begin{abstract}
Imbalanced image datasets are commonly available in the domain of biomedical image analysis. Biomedical images contain diversified features that are significant in predicting targeted diseases. Generative Adversarial Networks (GANs) are utilized to address the data limitation problem via the generation of synthetic images. Training challenges such as mode collapse, non-convergence, and instability degrade a GAN's performance in synthesizing diversified and high-quality images. In this work, MSG-SAGAN, an attention-guided multi-scale gradient GAN architecture is proposed to model the relationship between long-range dependencies of biomedical image features and improves the training performance using a flow of multi-scale gradients at multiple resolutions in the layers of generator and discriminator models. The intent is to reduce the impact of mode collapse and stabilize the training of GAN using an attention mechanism with multi-scale gradient learning for diversified X-ray image synthesis. Multi-scale Structural Similarity Index Measure (MS-SSIM) and Frechet Inception Distance (FID) are used to identify the occurrence of mode collapse and evaluate the diversity of synthetic images generated. The proposed architecture is compared with the multi-scale gradient GAN (MSG-GAN) to assess the diversity of generated synthetic images. Results indicate that the MSG-SAGAN outperforms MSG-GAN in synthesizing diversified images as evidenced by the MS-SSIM and FID scores.
\end{abstract}

\section{Introduction}
Generative adversarial networks (GANs) are generative models used for image synthesis in the computer vision domain \cite{wang2021generative}. GANs are composed of generator and discriminator models. The generator takes a random vector input and generates a noisy image. This image is passed to the discriminator model. The discriminator model classifies the generated images from the real images and provides gradient feedback to the generator. The generator model updates its learning of the feature distribution of real images through feedback provided by the discriminator. GANs work with adversarial training where the generator and the discriminator try to improve their performance based on each other's feedback \cite{goodfellow2020generative}.

GANs face difficulty in synthesizing images with complex and diverse features. This problem arises due to technical challenges that occur during the training of GANs. Training challenges include mode collapse, non-convergence, and instability \cite{jabbar2021survey}. Mode collapse refers to the generation of identical synthetic images by the generator regardless of diverse real images while the non-convergence and instability problem imbalanced the training due to the vanishing gradient problem. These problems limit the utility of GANs for image datasets with a diverse range of salient image features \cite{wu2021black}. In general, GANs are designed with convolutional neural networks (CNNs) that fail to capture image features such as texture, geometry, position, and color of the objects. One of the reasons could be that the CNNs mostly utilize convolutional features in modeling the dependencies over diverse image regions \cite{zhang2019self}.

In the domain of biomedical imaging, the diverse features of biomedical images are important to consider in disease recognition or computer-based diagnosis tasks \cite{liu2021survey}. These diverse features contain significant information about the disease being diagnosed and analyzed. GANs have been utilized for biomedical image synthesis. Several imaging modalities such as X-rays, Computed Tomography (CT), Magnetic Resonance (MR), Ultrasound, and Positron Emission Tomography (PET) have utilized GANs to generate synthetic samples \cite{alamir2022role}. The generation of diversified synthetic images is a significant barrier for GANs that limits their utility in the biomedical imaging domain. 

X-ray images are widely utilized to diagnose diseases in the human body. X-ray images contain a wide spectrum of disease features that help physicians to monitor diseases more accurately \cite{aggarwal2021diagnostic}. Publicly available X-ray image datasets are limited and imbalanced \cite{alvarez2022does}. Image synthesis is a potential means of augmenting and balancing these X-ray images. In image synthesis, synthetic images are produced by replicating the actual distributions of image features. Therefore, this method is significant as compared to the traditional augmentation approaches such as geometrical transformations \cite{shorten2019survey}. GANs have demonstrated remarkable advancements in image synthesis in the biomedical imaging domain \cite{ahmad2022new}.

State-of-the-art GANs such as ProGAN \cite{kim2021realistic}, StyleGAN \cite{hong20213d}, and MSG-GAN \cite{molahasani2022capsule} have been used for biomedical image synthesis. These GAN architectures have demonstrated significant performance in generating diverse images \cite{park2021realistic}. Minibatch discrimination, PixNorm, progressive growth of GAN layers, and Spectral normalization techniques have also been utilized to enhance the diversity of synthetic images. The multi-scale gradient technique enables the discriminator learning more robust for the classification of real and synthetic images \cite{karnewar2020msg}. Biomedical images contain salient disease features such as the location, size, color, and structure of the disease region of interest. These features are susceptible and important to predict and analysis of the disease. GANs learn images through convolutional features without giving attention to these salient features when generating synthetic images. However, it is important for a GAN to learn these biomedical image features during the training process.

In the domain of image recognition, self-attention is considered the best approach to focusing on diverse features of the images \cite{guo2022attention}. The self-attention measures relative information of features based on their feature maps and combines them globally with a weighted scoring function. Consequently, it helps to focus on the significant features for the specific application tasks \cite{zhang2019self}.

To address the training challenges of GANs, several GAN variants based on the attention mechanisms have attempted to improve the training performance of GANs for natural and biomedical images \cite{guo2022attention}. Self-attention improves the learning of generator and discriminator models in generating diversified biomedical images \cite{abdelhalim2021data}.

In order to balance and stabilize the training of a GAN, the loss function has also a great impact on the GAN's training performance for generating realistic synthetic images. Loss functions such as WGAN-GP, Hinge, and relativistic hinge losses have shown a reasonable improvement in generating diversified synthetic images \cite{jolicoeur2018relativistic}. However, the hinge loss has shown a great capacity to improve the GAN's learning to generate diverse biomedical images \cite{kim2021generative}.

The occurrence of mode collapse and diversity of synthetic images is assessed by the Multi-scale Structural Similarity Index Measure (MS-SSIM) and Frechet Inception Distance (FID). The MS-SSIM score can detect the lack of diversity using perceptual similarity measures in synthetic images while the FID score provides a distance between the feature distributions of real and synthetic images \cite{saad2022addressing}. 

This work contributes a novel GAN architecture for diversified X-ray image synthesis. The generator and discriminator models use multi-scale gradient learning to learn the gradient information at intermediate layers of the generator and discriminator models using multi-scale image resolutions during the training of GAN. A self-attention layer is proposed in the generator and discriminator models to learn the long-range dependencies of X-ray image features during training through a multi-scale gradient approach. The relativistic-hinge loss is used to stabilize the training and generate diverse synthetic images. The MS-SSIM and FID scores are used to evaluate the diversity of generated images. 
\begin{table}[hbt!]
\centering
\ra{1.1}
\caption{Attention mechanisms integrated into GANs for biomedical image analysis}
\begin{footnotesize}
      \resizebox{1\textwidth}{!}{%
      \begin{tabular}{lllllll}
      \toprule
      \textbf{Year} & \textbf{GAN\_Variant} & \textbf{Attention\_Type} & \textbf{Embedding} & \textbf{Image\_Type} & \textbf{Application\_Type} \\ \midrule
      2022 \cite{li2022explainable} & MtAA-NET & Multi-task Attention & Generator & CT & Segmentation \\
      2022 \cite{shang2022short} & CycleGAN & Channel Attention & Generator & PET & Reconstruction \\
      2021 \cite{tang2021plane} & AUGAN & Pixel-aware Attention & Generator & Ultrasound & Reconstruction \\
      2021 \cite{yin2021generative} & AMGAN & Dual Attention & Generator & MRIs & Segmentation \\
      2021 \cite{deng2021combining} & P2PGAN & Residual Attention & Generator & MRIs & Segmentation \\
      2021 \cite{abdelhalim2021data} & SPGGAN & Self Attention & Both & Dermoscopic & Synthesis \\
      2021 \cite{liu2021magan} & MAGAN & Mask Attention & Both & CT & Synthesis \\
      2020 \cite{kearney2020attention} & A-CycleGAN & Self Attention & Discriminator & MR-CT & Translation \\
      \bottomrule
      \end{tabular}}
      \end{footnotesize}
      \label{attngan}
\end{table}
\section{Related Work}
Several GAN models with modified architectures and loss functions have been proposed to improve the generation of diverse synthetic images. GAN architectures have been proposed with novel discriminators and generators based on the application domains. The performance of GANs has improved by embedding new convolutional layers, normalization, and regularization techniques in the generator and discriminator models \cite{radford2015unsupervised}\cite{miyato2018spectral}\cite{nie2020towards}. Several loss functions have been proposed to stabilize the training of GANs \cite{pan2020loss}. These advancements demonstrate significant improvements in GANs but have a limited scope for synthesizing improved diversified and high-quality images for different application domains.

In the domain of biomedical imaging, despite of above contributions, variants of attention mechanisms are proposed in GAN architectures to enhance the capacity of GANs to generate diversified and high-quality images as detailed in Table \ref{attngan}. Several attention mechanisms with GANs have been proposed for different applications such as image segmentation, image reconstruction, image synthesis, and image-image translation as detailed in Table \ref{attngan}. The attention mechanisms embedded in the generator, discriminator, or both models can improve the diversity and quality of generated images. These GANs utilize conditional information for the segmentation and reconstruction of biomedical images using different attention mechanisms. For image synthesis, self-attention with progressively growing GAN is proposed to generate diversified dermoscopic images. The authors succeed to alleviate partial mode collapse in their GAN architecture. Similarly, a mask-attention is proposed to generate high-quality Computed Tomography (CT) images with a conditional GAN. The authors utilize additional information on attention maps of targeted diseases to improve the quality of generated images. This approach also requires additional effort for mapping the attention masks of the diseases.

Generally, conditional masks of diseases are not available publicly in the domain of biomedical imaging. It requires an additional effort from physicians to annotate the disease masks. This problem limits the scope of GANs to only annotated biomedical image datasets. However, unconditional biomedical images require more work in the context of GANs to address this limitation. Therefore, this work investigates the utility of self-attention feature maps to guide a GAN using multi-scale gradient learning for synthesizing diversified biomedical images.     
\section{Methodology}
\begin{figure}
\includegraphics[width=1\textwidth,height=1\textheight,keepaspectratio]{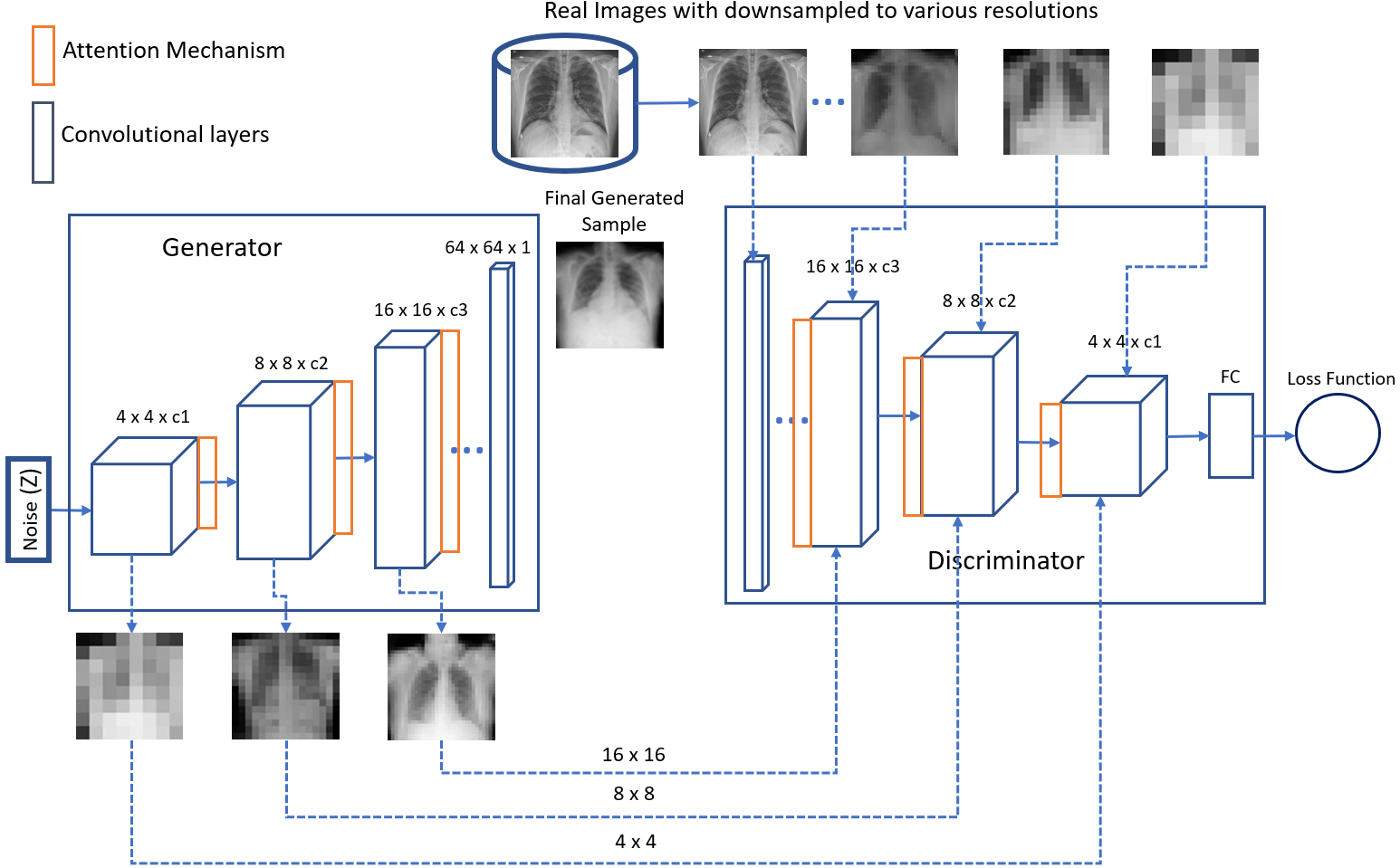}
\centering
\caption{The proposed architecture of MSG-SAGAN. The MSG-SAGAN is trained using multi-scale gradient learning at intermediate layers of the generator and discriminator models to generate X-ray images. The embedding of the self-attention mechanism in each block of the generator and discriminator models helps to generate improved diversified images through learning long-range dependencies of image features.} \label{SAMGAN}
\end{figure}

The workflow of the proposed approach has been depicted in Fig. \ref{SAMGAN}. The MSG-SAGAN generates synthetic X-ray images using multi-scale gradient learning between the intermediate layers of the generator and discriminator models. The generator and discriminator models are developed with the convolutional and self-attention layers to enable the relationships among long-range dependencies of image features for stabilizing the training and generating diversified X-ray images. Self-attention utilizes feature attention maps to improve the learning of the generator and discriminator models as depicted in Fig. \ref{SA}.
\subsection{Dataset}
In this work, the publicly available dataset of Corona Virus Disease (COVID-19) chest X-ray images is utilized \cite{rahman2021exploring}. The dataset contains 3616 X-ray images. The images were resized into 64x64 resolution. The X-ray images were preprocessed using a horizontal flipping to augment the data size.  
\subsection{GAN Architecture}
The Multi-scale Gradient Self-attention GAN (MSG-SAGAN) architecture utilizes a multi-scale gradient \cite{karnewar2020msg} learning approach between the generator and discriminator models. In MSG-SAGAN, the discriminator analyzes the output of the intermediate layers of the generator instead of looking only at the final layer output. The discriminator sends gradient feedback to multiple scales of the generator that helps a generator to create realistic diversified images. The training stabilizing techniques such as PixNorm and Mini-batch standard deviation are implemented within the GAN architecture. The PixNorm is embedded in the generator model to normalize the feature vectors. The Mini-batch standard layer is embedded into the discriminator of the GAN architecture to improve the diversity of generated image samples. The MSG-SAGAN architecture is trained with 500 epochs with a batch size of 16. As a baseline, the MSG-GAN \cite{karnewar2020msg} is reimplemented and trained on the CelebA dataset using the same parameters such as WGAN-GP loss, RMSprop optimizer, and 0.003 learning rates for the generator and discriminator models.

\begin{figure}
\centering
\includegraphics[width=1\textwidth,height=1\textheight,keepaspectratio]{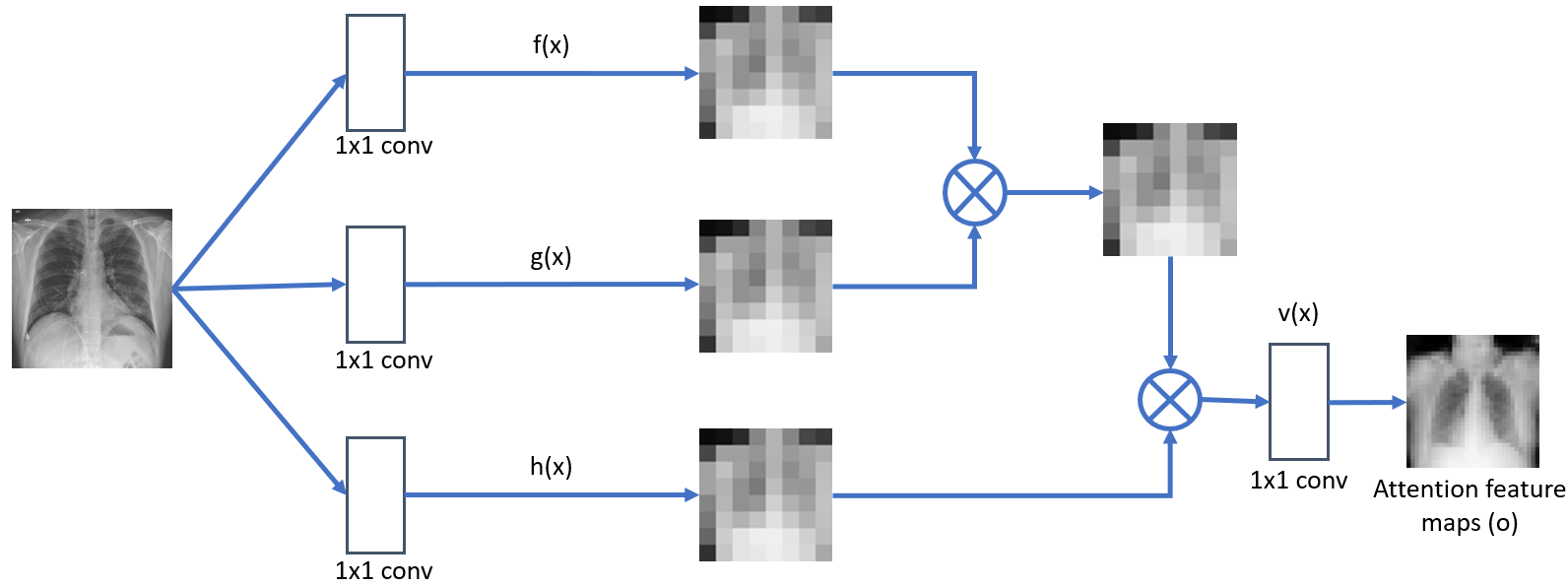}
\caption{Self-attention mechanism of MSG-SAGAN. The attention score is measured using different feature maps extracted from convolutional layers of the generator and discriminator models.} \label{SA}
\end{figure}
\subsubsection{Hyperparameters:}
The hyperparameters have a huge impact on the training performance of MSG-SAGAN architecture. The selection of efficient hyperparameters can improve the stability of GANs and their capacity to generate diversified synthetic images. In this work, the proposed MSG-SAGAN is trained with an Adam optimizer. The generator and discriminator models are fine-tuned using different learning rates such as 0.003, 0.0003, 0.0002, and 0.0001 to evaluate the MSG-SAGAN for diverse image synthesis. The equalized learning rates are used for both generator and discriminator models to balance the training of MSG-SAGAN.
\subsubsection{Spectral Normalization:}
Spectral normalization is used in the generator and discriminator models of the MSG-SAGAN. It helps the MSG-SAGAN avoid noisy gradients and enables fewer discriminator updates per generator, reducing the computational cost of training and improving the diversity of synthetic images.
\subsubsection{Loss Function:}
The experiments were conducted using a relativistic-hinge loss function as defined in Eq. \ref{Eq:Disloss} and \ref{Eq:Genloss}. Relativism in the hinge loss helps the discriminator to improve its learning by making predictions of the real images as half of the images are fake on average instead of taking them all as real. This prior training information helps the discriminator to classify and predict the real and fake images more accurately \cite{jolicoeur2018relativistic}. 
\begin{equation}
L_D^{\text {HingeGAN }}=\mathbb{E}_{x_r \sim \mathbb{P}}\left[\max \left(0,1-\tilde{D}\left(x_r\right)\right)\right]+\mathbb{E}_{x_g \sim \mathbb{Q}}\left[\max \left(0,1+\tilde{D}\left(x_g\right)\right)\right]\label{Eq:Disloss}
\end{equation}
\begin{equation}
L_G^{\text {HingeGAN }}=\mathbb{E}_{x_g \sim \mathbb{P}}\left[\max \left(0,1-\tilde{D}\left(x_g\right)\right)\right]+\mathbb{E}_{x_r \sim \mathbb{Q}}\left[\max \left(0,1+\tilde{D}\left(x_r\right)\right)\right]\label{Eq:Genloss}
\end{equation}
$$
\begin{aligned}
&\tilde{D}\left(x_r\right)=C\left(x_r\right)-\mathbb{E}_{x_g \sim \mathbb{Q}} C\left(x_g\right) \\
&\tilde{D}\left(x_g\right)=C\left(x_g\right)-\mathbb{E}_{x_r \sim \mathbb{P}} C\left(x_r\right)\label{Eq:G_and_D}
\end{aligned}
$$
In Eq. \ref{Eq:Disloss} and \ref{Eq:Genloss} as reported in \cite{jolicoeur2018relativistic}, discriminator and generator losses are defined for real and generated images. The real image samples are defined with $x_r$ and the generated samples are defined with $x_g$ where P and Q refer to the distributions of real and generated data respectively. The non-transformed layer is denoted by $C\left(x\right)$ while $D\left(x\right)$ denotes the transformed layer. 
\subsubsection{Self-Attention Mechanism:}
The self-attention is embedded in the generator and discriminator models of the MSG-SAGAN. The self-attention has a significant capacity for modeling relationships between diverse features in images. These diverse features include different spatial regions, channels, and pixels of images \cite{guo2022attention}. The self-attention utilizes two feature spaces $f$ and $g$ transformed by previous hidden layer $x \in$ $\mathbb{R}^{C \times N}$ to calculate the attention \cite{zhang2019self} shown in Fig. \ref{SA}. The attention function is calculated using the following equation where feature spaces $f$ and $g$ are $f(x)=W_f x, g(x)=W_g x$:
\begin{equation}
\beta_{j, i}=\frac{\exp \left(s_{i j}\right)}{\sum_{i=1}^N \exp \left(s_{i j}\right)}, \text { where } s_{i j}=f\left(x_i\right)^T g\left(x_j\right)\label{Eq:attention}
\end{equation}
In Eq. \ref{Eq:attention}, $\beta_{j, i}$ indicates the range of attention where the model computes mapping of $j^{\text {th}}$ location of the $j^{\text {th }}$ feature regions. Moreover, C denotes the number of channels while N denotes the number of feature locations of features transformed by the prior hidden layer. The output of the overall attention layer is formulated \cite{zhang2019self} as follows:
\begin{equation}
o_j=v\left(\sum_{i=1}^N \beta_{j, i} h\left(x_i\right)\right), h\left(x_i\right)=W_h x_i, v\left(x_i\right)=W_v x_i\label{attnoutput}
\end{equation}
In Eq. \ref{attnoutput}, the output $o$ is unrolled as $o=\left(o_1, o_2, \ldots, o_j, \ldots, o_N\right) \in$ $\mathbb{R}^{C \times N}$ while $W_g \in \mathbb{R}^{C \times C}, W_f \in \mathbb{R}^{C \times C}$, $\boldsymbol{W}_{\boldsymbol{h}} \in \mathbb{R}^{C \times C}$, and $\boldsymbol{W}_{\boldsymbol{v}} \in \mathbb{R}^{C \times C}$ are learned weight metrics. These weight metrics are implemented as 1x1 convolutions within the attention mechanism. The channel count is reduced as $c/k$ to improve the memory efficiency where $k$ is set to 8 as suggested in \cite{zhang2019self}.  

Furthermore, the output of the attention layer is multiplied by a scale parameter and appended back to the input feature map \cite{zhang2019self}. So, the final output of the self-attention layer will be:
\begin{equation}
y_i=\gamma o_i+x_i\label{attnfinal}
\end{equation}
In Eq. \ref{attnfinal}, $\gamma$ is a learnable scale parameter that is initialized at zero.

\subsection{Identification of Mode Collapse Problem}
The occurrence of mode collapse is identified by the MS-SSIM. The MS-SSIM computes the similarity score between two images using contrast, structure, and luminance features. MS-SSIM score is measured using randomly selected image pairs from the dataset to asses the diversity of synthetic images. The diversity of images is compared by measuring the MS-SSIM score from the real dataset and synthetic image dataset generated by GANs. A higher MS-SSIM score of the synthetic dataset indicates the occurrence of mode collapse in GANs. MS-SSIM can be computed between two image samples $a$ and $b$ as defined in Eq. \ref{Eq. ms-ssim} \cite{borji2019pros}.
\begin{equation}
\resizebox{0.75\hsize}{!}{$%
\operatorname{MS}-\operatorname{SSIM}(a, b)=I_{M}(a, b)^{\alpha_{M}} \prod_{j=1}^{M} C_{j}(a, b)^{\beta_{j}} S_{j}(a, b)^{\gamma_{j}}\label{Eq. ms-ssim}
$}%     
\end{equation}
Contrast (C) and structural (S) features of images are computed at scale $j$ as denoted in Eq. \ref{Eq. ms-ssim}. Luminance (I) is calculated at the coarsest scale (M). The $\alpha$, $\beta$, and $\gamma$ are the weight parameters as detailed in \cite{wang2003multiscale}. In this work, 3616 real and 3616 synthetic X-ray images are used to compute the MS-SSSIM scores of real and synthetic image datasets.

\subsection{Evaluation of the Diversity and Quality of Synthetic X-ray Images}
The diversity and quality of generated images are evaluated using the FID scores. FID computes the Wasserstein-2 distance between synthetic images and real images using feature activations \cite{miyato2018cgans}. It captures the multivariate Gaussian activations by calculating the mean and covariance of the images (real and synthetic) using the last pooling layer of an Inception-V3 model. The FID score is calculated as shown in Eq. \ref{FID}, \cite{borji2019pros}.
\begin{equation}
\resizebox{0.75\hsize}{!}{$%
FID(r, s)=\left\|\mu_{r}-\mu_{s}\right\|_{2}^{2}+\operatorname{Tr}\left(\Sigma_{r}+\Sigma_{s}-2\left(\Sigma_{r} \Sigma_{s}\right)^{\frac{1}{2}}\right)\label{FID}
$}%
\end{equation}
In Eq. \ref{FID}, $r$ and $s$ denote real and synthetic images while $\left(\mu_{r}, \Sigma_{r}\right)$ and $\left(\mu_{s}, \Sigma_{s}\right)$ denote the mean and covariances of real and synthetic images. The FID score ranges from 0.0 to $+\infty$. The higher FID score shows a larger distance between synthetic and real data distributions that indicates the occurrence of mode collapse \cite{borji2019pros}. A lower FID score shows a smaller distance between synthetic and real data distributions that indicates a higher degree of diversity. This work measures FID using 3616 real and 3616 generated images.
\begin{table}[hbt!]
\centering
\ra{1.1}
\caption{Analysis of the MS-SSIM and FID scores for the proposed MSG-SAGAN architecture and the MSG-GAN architecture to evaluate the diversity of generated synthetic X-ray images. Best scores are highlighted in bold values.}
\begin{footnotesize}
      \resizebox{1\textwidth}{!}{%
      \begin{tabular}{lllllllllllll}
      \toprule
      \textbf{GANs} & \textbf{PN} & \textbf{SN} & \textbf{MBD} & \textbf{AM} & \textbf{FA} & \textbf{Opt} & \textbf{LR} & \textbf{Loss} & \textbf{Data} & \textbf{FID} & \textbf{MR} & \textbf{MG} \\ \midrule
      MSG-GAN \cite{karnewar2020msg} & \checkmark & x & \checkmark & x & \checkmark & RMSprop & 0.003 & WGAN-GP & CelebA & 8.86 & - & - \\
      MSG-GAN$_{1}$ (Re) & \checkmark & x & \checkmark & x & \checkmark & RMSprop & 0.003 & WGAN-GP & CelebA & 18.5 & - & -  \\
      MSG-GAN$_{2}$ & \checkmark & x & \checkmark & x & \checkmark & RMSprop & 0.003 & WGAN-GP & X-ray & 380 & 0.50 & 0.74 \\
      MSG-GAN$_{3}$ & \checkmark & x & \checkmark & x & x & RMSprop & 0.003 & WGAN-GP & X-ray & 310.2 & 0.50 & 0.51 \\
      MSG-GAN$_{4}$ & \checkmark & x & \checkmark & x & \checkmark & Adam & 0.003 & WGAN-GP & X-ray & 330.33 & 0.50 & 0.66 \\
      MSG-GAN$_{5}$ & \checkmark & x & \checkmark & x & \checkmark & RMSprop & 0.003 & RLHinge & X-ray & 200 & 0.50 & 0.50 \\
      MSG-GAN$_{6}$ & \checkmark & x & \checkmark & x & \checkmark & Adam & \textbf{0.003} & \textbf{RLHinge} & X-ray & \textbf{167.1} & 0.50 & \textbf{0.47} \\
      MSG-GAN$_{7}$ & \checkmark & x & \checkmark & x & x & Adam & 0.003 & RLHinge & X-ray & 194.85 & 0.50 & 0.51 \\
      MSG-SAGAN$_{1}$ & \checkmark & x & \checkmark & \checkmark & x & Adam & 0.003 & RLHinge & X-ray & 557.28 & 0.50 & 0.99 \\
      MSG-GAN$_{8}$ & \checkmark & x & \checkmark & x & x & Adam & 0.0003 & RLHinge & X-ray & 217.3 & 0.50 & 0.54 \\
      MSG-GAN$_{9}$ & \checkmark & x & \checkmark & x & x & Adam & 0.0002 & RLHinge & X-ray & 272.02 & 0.50 & 0.53 \\
      MSG-GAN$_{10}$ & \checkmark & x & \checkmark & x & x & Adam & 0.0001 & RLHinge & X-ray & 254.0 & 0.50 & 0.58 \\
      MSG-GAN$_{11}$ & \checkmark & \checkmark & \checkmark & x & x & Adam & 0.003 & RLHinge & X-ray & 413.8 & 0.50 & 1.0 \\
      MSG-SAGAN$_{2}$ & \checkmark & \checkmark & \checkmark & \checkmark & x & Adam & 0.003 & RLHinge & X-ray & 413.8 & 0.50 & 1.0 \\
      MSG-SAGAN$_{3}$ & \checkmark & \checkmark & \checkmark & \checkmark & x & Adam & 0.0003 & RLHinge & X-ray & 387.2 & 0.50 & 0.55 \\
      MSG-SAGAN$_{4}$ & \checkmark & \checkmark & \checkmark & \checkmark & x & Adam & 0.0002 & RLHinge & X-ray & 198.8 & 0.50 & 0.55 \\
      MSG-SAGAN$_{5}$ & \checkmark & \checkmark & \checkmark & \checkmark & x & Adam & 0.0001 & RLHinge & X-ray & 282.2 & 0.50 & 0.54 \\
      MSG-SAGAN$_{6}$ & \checkmark & \checkmark & \checkmark & \checkmark & \checkmark & Adam & 0.0003 & RLHinge & X-ray & 243.0 & 0.50 & 0.47 \\
      MSG-SAGAN$_{7}$ & \checkmark & \checkmark & \checkmark & \checkmark & \checkmark & Adam & 0.0002 & RLHinge & X-ray & 366.2 & 0.50 & 0.53 \\
      MSG-SAGAN$_{8}$ & \checkmark & \checkmark & \checkmark & \checkmark & \checkmark & Adam & \textbf{0.0001} & \textbf{RLHinge} & X-ray & \textbf{139.6} & 0.50 & \textbf{0.50} \\
      \bottomrule
      \multicolumn{13}{l}{Re: Reimplemented; PN: Pixel Norm; SN: Spectral Norm; MBD: Minibatch Std Dev} \\
      \multicolumn{13}{l}{AM: Attention Mechanism; FA: Flip Augment; Opt: Optimizer; LR: Learning Rate} \\
      \multicolumn{13}{l}{MR: MS-SSIM Real; MG: MS-SSIM Generated; RLHinge: Relativistic Hinge}
      \end{tabular}}
      \end{footnotesize}
      \label{eval_score}
\end{table}
\section{Results and Discussion}
The MSG-SAGAN is proposed to alleviate the mode collapse in the MSG-GAN and improve the diversity of generated synthetic images in the context of X-ray images. MSG-SAGAN is a variant of MSG-GAN that utilizes an attention mechanism with multi-scale gradient learning to enhance the efficacy of synthesizing improved diversified X-ray images. The MS-SSIM score is used to identify the occurrence of mode collapse while the FID scores are used for the evaluation of the diversity in synthetic images. Resultant MS-SSIM and FID scores of MSG-GAN and MSG-SAGAN architectures are compared under a range of parameter settings as denoted in Table \ref{eval_score}.   

The reimplementation of the MSG-GAN as detailed in \cite{karnewar2020msg} resulted in a higher FID score than the original work when evaluated against the CelebA dataset. This was likely due to the number of real and synthetic images used in the calculation of FID. These details are omitted from \cite{karnewar2020msg} while in this work 10,000 real and 10,000 synthetic images were used in calculating the FID.

In the context of diverse synthetic X-ray images, the MSG-GAN$_{2}$ is trained using the same parameter settings including the loss, optimizer, learning rate, and horizontal flipping data augmentation. MSG-GAN underperformed in synthesizing diversified X-ray images as indicated by the degraded MS-SSIM and FID scores. 

The WGAN-GP loss is used to stabilize the training of GANs by avoiding the vanishing gradient problem. However, the RMSprop optimizer does not converge the training using the WGAN-GP loss for X-ray images because the RMSprop only relies on the second-order moment of gradients which leads to unstable training. Therefore, this parameter setting of MSG-GAN was not efficient to alleviate the mode collapse, stabilize the training, and generate diversified X-ray images. 

The X-ray images contain salient features such as the spine, heart, and lungs with their visual signatures like ribs, aortic arch, and distinct curvature of lower lungs. All these features are important to learn by the discriminator so that it can provide constructive feedback to the generator model. So, a GAN should focus on these X-ray image features when generating synthetic images. The proposed architecture of MSG-SAGAN has the capacity to learn these X-ray features using the attention feature maps as depicted in Fig. \ref{SA}.    

Firstly, the effect of data augmentation is analyzed. The MSG-GAN$_{3}$ does not utilize the horizontal flipping and the results of MS-SSIM and FID are slightly improved but no significant improvement was seen as the higher MS-SSIM score of synthetic X-ray images than the MS-SSIM score of real images indicates the occurrence of mode collapse. The MSG-GAN$_{7}$ with Adam optimizer and relativistic hinge loss is trained without horizontal flipping but the results were degraded as compared to MSG-GAN$_{6}$ with flipping. Furthermore, the MSG-SAGAN$_{6-8}$ utilizes the horizontal flipping and alleviated mode collapse, and improved the diversity of synthetic images as compared to the MSG-SAGAN$_{3-5}$ that does not utilize the horizontal flipping.  

Secondly, the MSG-GAN$_{4}$ is trained with an Adam optimizer and WGAN-GP loss that degrade the results. Moreover, the MSG-GAN$_{6-11}$ and MSG-SAGAN$_{1-8}$ are trained with the Adam optimizer and the relativistic hinge loss that alleviates the mode collapse and improves the diversity of generated images. The degraded results are evident from the other parameters such as spectral norm and attention mechanism. The Adam optimizer outperformed RMSprop due to the fact that it has the capacity to stabilize the training and converge faster because it uses both first and second-order moments of the gradients.

Thirdly, the relativistic hinge loss is used with the Adam and RMSprop optimizer in the MSG-GAN$_{6-11}$ and MSG-SAGAN$_{1-8}$. The relativistic hinge loss indicates significant improvement to alleviate the mode collapse and improve the diversity of synthetic images because relativism in the hinge loss helps a discriminator to provide constructive feedback to the generator.

The learning rate has a huge impact on the training of the GAN architectures. The most performant learning rate for MSG-GAN was 0.003 while 0.0001 for MSG-SAGAN. This happens because the multi-scale gradient learning stabilizes the training with a learning rate of 0.003 while the self-attention mechanism balances the training with a learning rate of 0.0001 as indicated in Table \ref{eval_score}.

Results indicate that spectral normalization degrades the training of the MSG-GAN while improving the training of the MSG-SAGAN as indicated in Table \ref{eval_score}. In the MSG-GAN, spectral normalization degrades the significant gradients that are flowing between the generator and the discriminator models (See MSG-GAN$_{11}$). Whereas, spectral normalization helps to avoid noisy gradients that are produced during the training of MSG-SAGAN due to the attention mechanism. 

MSG-SAGAN$_{8}$ outperforms the MSG-GAN$_{6}$ in terms of synthesizing diversified images and stabilizing the training process. Integrating the self-attention mechanism improves the flow of multi-scale gradients between the generator and discriminator models with small learning rates while degrading with large ones. The multi-scale gradients help improve the generator's learning capacity and discriminator models by propagating the gradients between the intermediate layers of the generator to the discriminator and vice versa. Consequently, the feature attention maps help a GAN to make relationships between long-range dependencies of the diverse image features.

The most performant MSG-GAN$_{6}$ instance results in an improved MS-SSIM of 0.474 for synthetic X-ray images as compared to real images and an FID of 167.1. However, the most performant MSG-SAGAN$_{8}$ instance results in an improved MS-SSIM of 0.50 for synthetic X-ray images as compared to real images and an improved FID of 139.6. The MS-SSIM and FID scores for MSG-SAGAN$_{8}$ indicate a stable training period and a reduction in the impact of mode collapse while synthesizing improved diversified X-ray images as compared to the alternate instances evaluated. 
\section{Conclusion}
In this work, MSG-SAGAN was proposed to reduce the impact of mode collapse and training instability for generating synthetic X-ray images. The MSG-SAGAN demonstrated an improved capacity for the synthesis of diversified X-ray images using the attention mechanism as compared to the MSG-GAN. The MSG-SAGAN was evaluated under different settings to quantify their impact on the diversity of synthetic images generated. Results were evaluated using the MS-SSIM and FID scores. The most performant MS-SSIM (0.50) and FID (139.6) were produced by MSG-SAGAN.

The MS-SSIM and FID scores indicate that the multi-scale gradients approach in a GAN is performant with a learning rate of 0.003 for X-ray images. However, an attention mechanism with multi-scale gradient learning is the most performant with a learning rate of 0.0001. These results of MS-SSIM and FID demonstrate the impact of learning rates in the training of GANs to synthesize diversified X-ray images. A learning rate of 0.0001 utilizes small training steps to update the gradient weights for each iteration to converge the MSG-SAGAN training to balance and stabilized training. 

Spectral normalization degrades the training stability of MSG-GAN while improving the training stability of MSG-SAGAN. Adam was the most performant optimizer in both MSG-GAN and MSG-SAGAN. Relativistic hinge loss stabilizes the training and improves the generation of diversified X-ray images. The data augmentation of horizontal flipping indicates a significant improvement in stabilizing the training of MSG-SAGAN to synthesize diversified X-ray images. Horizontal flipping provides mirror copies of X-ray images that improve the learning of MSG-SAGAN with training more on salient features of X-ray images.   

In future work, different variants of attention mechanisms will be investigated with a multi-scale gradient approach in the GAN architecture for synthesizing X-ray images. The self-attention will be integrated with different positions in the generator and discriminator models or only with the generator or discriminator model in the MSG-SAGAN. Different learning rates will also be investigated to synthesize the improved diversified X-ray images. This work will be extended with the integration of self-attention and its variants into state-of-the-art GANs such as StyleGAN V3, and Projected GANs.

% \bibliographystyle{basic}
% \bibliography{References.bib}

\end{document}